\documentstyle[epsf,preprint,aps]{revtex}
\begin{document}
\def\etal{{\it et.~al.}}
\def\en{\langle n\rangle}
\def\vg{{\bf G}}
\def\vk{{\bf k}}
\def\vkp{{\bf k'}}
\def\vp{{\bf p}}
\def\vpp{{\bf p'}}
\def\vq{{\bf q}}
\def\ubar{\bar{U}}
\def\ep{\epsilon_\vp}
\def\epq{\epsilon_{\vp+\vq}}
\def\DelU{\Delta_U}
\def\Delk{\Delta_\vk}
\def\Delp{\Delta_\vp}
\def\Delpp{\Delta_\vpp}
\def\Delpq{\Delta_{\vp+\vq}}
\def\Ek{E_{\vk}}
\def\Ekp{E_{\vkp}}
\def\Ep{E_{\vp}}
\def\Epp{E_{\vpp}}
\def\Epq{E_{\vp+\vq}}
\def\cos{\mbox{cos}}
\def\iv{\mbox{Im} V}
\def\taut{\tau^{-1}(\vp,\omega)}
\def\taun{\tau^{-1}_N(\vp,\omega)}
\def\tauu{\tau^{-1}_U(\vp,\omega)}
\def\dwave{d_{x^2-y^2}}
\def\2del{2\Delta_0/k_BT_c}

%
%
%

\title{Quasi-particle Lifetimes in a \boldmath${d_{x^2-y^2}}$ Superconductor}

\author{Daniel Duffy\thanks{duffy@erdc.hpc.mil}} 
\address{Computer Sciences Corporation,
2545 South I-20 Frontage Road, Suite A,
P.O. Box 820186,
Vicksburg, MS 39182}

\author{P.J. Hirschfeld\thanks{pjh@phys.ufl.edu}}

\address{Department of Physics, University of Florida, 
Gainesville, FL 32611}

\author{Douglas J.~Scalapino\thanks{djs@vulcan.physics.ucsb.edu}}

\address{Department of Physics, University of California, 
Santa Barbara CA 93106}

\date{\today}
\maketitle

\begin{abstract}
\baselineskip .5cm
We consider the lifetime of quasi-particles in a d-wave superconductor
due to scattering from antiferromagnetic spin-fluctuations, and
explicitly separate the contribution from Umklapp processes which
determines the electrical conductivity.  Results for the temperature
dependence of the total scattering rate and the Umklapp scattering
rate are compared with relaxation rates obtained from thermal and
microwave conductivity measurements, respectively.
\end{abstract}

\pacs{PACS numbers: }
\narrowtext

It is now widely accepted that the superconducting state of the high $T_c$
cuprates is characterized by a $d_{x^2-y^2}$-order parameter\cite{DJS95,DvH95}.
 Furthermore,
various transport properties such as NMR relaxation times\cite{BS91,TPL93}
 and the microwave
conductivity\cite{HPS94} 
have been calculated within a quasi-particle BCS-RPA framework
which phenomenologically takes into account the interplay of both
the $d_{x^2-y^2}$ pairing and the antiferromagnetic spin correlations. In
particular, this
approach has previously provided reasonable fits
of the temperature
dependence of the  NMR $T_1$ and $T_2$ relaxation times in optimally
doped YBa$_2$Cu$_3$O$_{7-\delta}$\cite{Martindale,Kitaoka}.
 With recent advances in the preparation of YBCO crystals
\cite{EWF96,LBH98},
microwave measurements \cite{Hos99} at a variety of higher frequencies, and 
$\kappa_{xy}$
thermal conductivity measurements \cite{Zhapp}, one has new information on the
quasi-particle lifetime involved in the electrical conductivity and
the mean-free path associated with the
thermal conductivity.  Several puzzles have presented themselves.
First, microwave measurements appear to indicate that the relaxation
rate, long known to collapse rapidly below the critical 
temperature\cite{Nus91,Rom92,Bon92}, varies as $T^4$, or possibly
exponentially, at temperatures $T\ll T_c$.  On the
other hand, the  theory of
the quasi-particle relaxation rate 
in a $d$-wave superconductor due to electron-electron interactions predicts a
$T^3$ variation above a scale set by impurity 
scattering\cite{BS91,QSB94,Yas95}.  
A solution to this discrepancy has recently been suggested by
Walker and Smith\cite{WS00}, who noted that in the absence of
disorder, the transport lifetime entering the microwave conductivity
arises solely from Umklapp processes, which are always gapped below
$T_c$ even in  
a $d$-wave superconductor, leading to
an exponential decay $1/\tau_{U}\sim \exp(-\Delta_U/T)$. 
Secondly, while early thermal conductivity experiments extracted 
quasi-particle mean-free paths in apparent agreement with 
microwave measurements, recent measurements indicate that 
on cleaner crystals, where the mean-free path due to inelastic
scattering can be probed to lower temperatures, the extracted 
mean-free path is shorter than in the microwave case\cite{Zhapp}.
Furthermore, the inelastic contribution to the inverse thermal mean 
free path appears to follow a $T^3$ temperature dependence at low
temperatures (see below).
\vskip .2cm
Motivated by the proposal of Walker and Smith\cite{WS00}
regarding the role of Umklapp processes, we investigate whether
the same type of spin-fluctuation interaction used to fit
the NMR data can also provide a quantitative fit to the lifetimes
extracted from the microwave and thermal conductivities.  In particular,
we wish to understand if
the discrepancies discussed above can be resolved by
recognizing that normal electron-electron interactions conserve 
the total momentum
and therefore cannot degrade the electric current, whereas 
they can degrade the heat current \cite{DL99}.   
We therefore calculate, within a BCS model in which the 
spin-fluctuation interaction is accounted for in the random 
phase approximation (RPA),
the total quasi-particle lifetime and the Umklapp-only
quasi-particle lifetime and compare these with the $\kappa_{xy}$ thermal
conductivity and microwave conductivity measurements, respectively. We find
that the inelastic 
lifetimes extracted from thermal transport and charge transport
can be accounted for within a BCS-RPA quasi-particle framework.
Furthermore, the parameter values which are required are consistent
with what is known
about the Fermi surface\cite{Chi99,Sch97,Sch98} and 
spin-fluctuations from other measurements\cite{Martindale,Kitaoka}.
The analysis is based, however, on the subtraction of a  
constant relaxation rate from both quantities, an approach which
must be justified within microscopic theory; we present a brief
discussion of these issues and the light shed by our analysis on
the behavior of the low-energy impurity scattering amplitude.


In the following, we will assume that we have BCS quasi-particles with
a $\dwave$ gap
\begin{equation}
\Delta_{\bf k}(T) = \frac{1}{2}\Delta_0(T)(\cos k_x - \cos k_y)
\end{equation}
moving on a 2D square lattice with near- and next-near-neighbor
one-electron hopping matrix elements $t$ and $t'$, respectively.  The
effective electron-electron interaction will be modeled by the
spin-fluctuation form 
\begin{equation}
\label{eq:V}
\iv(\vq,\omega) = \frac{3}{2}\ubar^2
\frac{\chi_0''(\vq,\omega)}{(1-\ubar\chi_0'(\vq,\omega))^2
+(\ubar\chi_0''(\vq,\omega))^2}.
\end{equation}
Here, $\ubar$ and the chemical potential, $\mu$, which sets the
average site occupation, $\en$, control the strength of the
antiferromagnetic spin-fluctuations, and $\chi_0(\vq,\omega)$ is the
BCS susceptibility given by
\begin{eqnarray}
\label{eq:chi}
\chi_0(\vq,\omega) = \frac{1}{N}\sum_{\vp} \mbox{\huge{\{}} 
 &\frac{1}{2}& \left[1+\frac{\epq\ep+\Delpq\Delp}{\Epq\Ep}\right]
\frac{f(\Epq)-f(\Ep)}{\omega-(\Epq-\Ep)+i0^+} \nonumber \\
+&\frac{1}{4}& \left[1-\frac{\epq\ep+\Delpq\Delp}{\Epq\Ep}\right]
\frac{1-f(\Epq)-f(\Ep)}{\omega+(\Epq+\Ep)+i0^+}\\
+&\frac{1}{4}& \left[1-\frac{\epq\ep+\Delpq\Delp}{\Epq\Ep}\right] 
\frac{f(\Epq)+f(\Ep)-1}{\omega-(\Epq+\Ep)+i0^+}\mbox{\huge{\}}},\nonumber
\end{eqnarray}
where $\Ep=\sqrt{\ep^2+\Delp^2}$ and
$\ep = -2t(\cos p_x+\cos p_y)-4t'\cos p_x\cos p_y-\mu$.
The band parameters are taken to be  $t'/t=-0.35$ with a site filling
of $\langle n_{i,\uparrow}+n_{i,\downarrow}\rangle=0.85$, and
the temperature dependence of the gap has been parametrized as
\begin{equation}
\label{eq:gap}
\Delta_0(T) = \frac{\Delta_0}{2}\mbox{tanh}(\alpha\sqrt{T_c/T-1}).
\end{equation}
We have used $\alpha=3.0$ to reproduce the rapid onset of the gap
below $T_c$ which is observed experimentally, and 
set $\2del=5$.  Throughout this paper, we  use
$t=1$ as our energy scale and  take $\ubar/t=2.2$; this is similar to the
$\ubar=2.0$
used in previous 
calculations \cite{BS91} of the NMR $T_1$ and $T_2$ relaxation
rates, where   an RPA form for $\chi(\vq,\omega)$ similar to
Eq.~(\ref{eq:chi}) was employed, but  a simple ($t'=0$) tight-binding band
assumed.

Using the effective interaction, Eq.~(\ref{eq:V}), the Fermi Golden
Rule expression for the quasi-particle lifetime is \cite{QSB94}
\begin{eqnarray}
\label{eq:tau}
\taut = &\int& \frac{d^2p'}{(2\pi)^2}
\iv(\vp-\vpp,\omega-\Epp)(1+\frac{\Delp\Delpp}{\omega\Epp})
[n(\omega-\Epp)+1][1-f(\Epp)] \\\nonumber
+ &\int& \frac{d^2p'}{(2\pi)^2}
\iv(\vp-\vpp,\omega+\Epp)(1-\frac{\Delp\Delpp}{\omega\Epp})
[n(\omega+\Epp)+1]f(\Epp),
\end{eqnarray}
where $n(\omega)$ and $f(\omega)$ are the usual Bose and Fermi
factors.  Here the first term arises from quasi-particle scattering
resulting from the absorption and emission of a spin-fluctuation.
The second term corresponds to a process in which a
quasi-particle recombines with another quasi-particle to form a pair
and the excess energy is emitted as a spin-fluctuation.

An example of the type of process which this includes is
schematically illustrated in Fig.~1(a).   Here, a quasi-particle with 
momentum
$\vp$ scatters off of another quasi-particle with momentum $\vk$ leading
to an intermediate state containing quasi-particles of momentum $\vpp$ 
and $\vkp$.  Momentum conservation on the lattice implies that
\begin{equation}
\label{eq:energy}
\vp+\vk=\vpp+\vkp+\vg
\end{equation}
where $\vg$ is a reciprocal lattice vector and all other momenta are
confined to the first Brillouin zone.  Since $\Ekp =
E_{\vp-\vpp+\vk-\vg} = E_{\vp-\vpp+\vk}$, we have the expression for
$\chi_0(\vq,\omega)$ given by Eq.~(\ref{eq:chi}) with $\vq=\vp-\vpp$
and with the $\vk$ sum extending over the entire first Brillouin zone.
If, however, we had not allowed any reciprocal lattice vectors except
$\vg=(0,0)$, the $\vk$ sum would, for given values of $\vp$ and
$\vpp$, extend only over that part of the first Brillouin zone for
which $\vkp=\vp-\vpp+\vk$ remains inside the first Brillouin zone.
This restriction of $\vg=(0,0)$ eliminates all of the Umklapp
scattering processes, allowing only ``normal'' processes.  In this
way, one can introduce a ``normal'' scattering lifetime $\taun$.  This
is obtained by using an interaction with the same denominator as in
Eq.~(\ref{eq:V}), but with $\chi_0''(\vq,\omega)$ in the numerator
replaced by the imaginary part of Eq.~(\ref{eq:chi}) with
$\vq=\vp-\vpp$ for which the sum over $\vk$ is restricted such that
$\vp-\vpp+\vk$ remains in the first Brillouin zone.  With this
restriction, the resulting ``normal'', non-Umklapp, part of
$\chi_0''(\vq,\omega)$ depends separately on $\vp$ and $\vpp$ rather
than only on the difference of $\vp-\vpp$.  Having calculated both
$\taut$ and $\taun$, one can obtain a lifetime which contains only the
Umklapp processes by subtraction
\begin{equation}
\tauu = \taut-\taun.
\label{seven}
\end{equation}

At temperatures which are  small compared with the maximum gap, the thermally
excited quasi-particles occupy states close to the $\dwave$ gap nodes
on the Fermi surface.  When two of these thermal quasi-particles
scatter, the outgoing states must also occupy states near the nodal
region in order to conserve energy.  For the Fermi surface shown in
Fig.~1(b), these scattering processes must have $\vg=(0,0)$ and only
contribute to the ``normal'' processes.  Umklapp scattering processes
are also possible, but 
these processes are exponentially suppressed at
low temperatures.  Following Walker and Smith\cite{WS00}, the reason for this is
illustrated in Fig.~1(b).  Here, we have set the initial quasi-particle
momentum $\vp$ to a node and numerically searched for the lowest
energy quasi-particle state $\vk$ such that momentum is conserved as in
Eq.~(\ref{eq:energy}) with the constraint that $\vg=(0,0)$.  For
$t'/t=-0.35$ and $\en=0.85$ corresponding to the Fermi surface shown,
the momentum $\vk$ is the lowest energy state for which
a quasi-particle can have an Umklapp scattering with the nodal
quasi-particle $\vp$. Since $\vk$ has moved away from the nodal region,
$\Delta_k$ is now finite.
When $k_BT$ is less than $\Delk$, the probability of finding such a
quasi-particle varies as $e^{-\Delk/k_BT}$ and the Umklapp
scattering is suppressed for
temperatures below $\Delk$.  We will call this special
value the Umklapp gap, $\DelU$. It depends upon the shape of the
Fermi surface which is set by $t^\prime/t$ and $\langle n\rangle$.  
Figure 2 shows the  Umklapp gap as a
function of the strength of the $t'/t$ for a fixed filling of
$\en=0.85$.  It is easily seen that an Umklapp gap exists even when
$t'/t=0$ and has a magnitude of about 0.07$t$ for the
band with $t'/t=-0.35$.  We note further that this corresponds
to a ``Dirac cone anisotropy ratio" $v_F/v_2\simeq 12$, which
is consistent with the value determined by low temperature
thermal conductivity measurements \cite{Chi99}. 

We now compare  the absolute values of the relaxation times in question.
In Fig.~3(a), we have plotted $\hbar/k_BT_c\tau_U$
and $\hbar/k_B T_c\tau$ versus $T/T_c$ for the parameters discussed above. 
In order to
compare recent thermal conductivity measurements with the total scattering
rate, we need to specify a Fermi velocity.  For the $t$-$t^\prime$
dispersion relation one has $v_F \simeq 2.16ta$ and taking
$T_c=0.1t=90K$, this gives $v_F \sim 10^7$cm/sec. Zhang \etal~\cite{Zhapp} 
have used an
effective Fermi velocity $\eta v_f \sim 10^7$ cm/sec with $\eta=0.6$ and
$v_F=1.7\times 10^7$ cm/sec.
Alternatively, Chiao \etal~\cite{Chi99}
have suggested that $v_F =2.5\times 10^7$ cm/sec. In the fit of
$\hbar/k_BT_c\tau(T)$ shown as the solid line in Fig.~4(a), we have used
$v_F=3\times10^7$ cm/sec to convert the thermal mean-free path data of
Ref.~\cite{Zhapp} to a quasi-particle relaxation rate shown as the 
open symbols.
Beyond the
choice of $v_F$, the determination of the 
quasi-particle lifetime from the thermal conductivity measurements is
complicated by the phonon contribution to the heat current, and necessitates 
certain assumptions regarding the nature of
quasi-particle scattering in the vortex state\cite{Zhapp}
which are still controversial.
However, we note that the magnitude of the theoretically
determined total scattering 
rate of $\hbar/\tau (T_c)\sim 3k_BT_c$ at the node is in 
rough agreement with that determined
by ARPES \cite{Kam00}, while $\hbar/\tau_U(T_c) \sim k_BT_c$ is consistent with
optical data.  In Fig 3(b), we show the low temperature range of
the theoretical predictions, along
with the Hosseini \etal \cite{Hos99} $1/\tau$ extracted from microwave
data using a two-fluid analysis, shown as open diamonds. Here one sees
that taking
$\bar U/t=2$ provides a reasonable fit to the magnitude of the low
temperature microwave scattering lifetime and the exponential decay of
the Umklapp processes also appears to provide a sensible explanation for
the observed temperature dependence at low temperatures. 


Clearly, there are a number of parameters in this calculation, and the
results do vary with their values.  Here, we have selected values for
$t'/t$ and the filling $\en$ to give a Fermi surface which resembles
that seen in ARPES data and various band structure calculations \cite{Sch97}
for the $d_{x^2-y^2}-p\sigma$ band of
YBCO.  Specifically, the Fermi surface, as seen in Fig.~1(b), has a
diagonal nodal k-spacing of $(0.8,0.8)\pi/2a$, and fixes the
size of the Umklapp gap, $\DelU$, which determines how
$\tauu$ decreases below the total scattering lifetime.
Note that for the
Fermi surface pictured in Fig.~1(b), $\2del=5$ corresponds to a
maximum of the gap on the Fermi surface near $(\pi,0)$ of
$\mbox{max}[2\Delta_0(\vp)/k_BT_c]=4.7$.
The choice of $\ubar/t=2.2$ was made consistent with
Refs. \cite{BS91,TPL93} and because it provided a reasonable fit to the
absolute magnitude of the microwave scattering lifetime, as discussed
below.

It is  important to ask the extent to which the data
support the fundamental prediction of this work, 
that the relaxation rate entering the thermal conductivity
data should be the total quasi-particle relaxation rate, which
varies as $(T/T_c)^3$ due to spin-fluctuation scattering and
recombination processes, in contrast to the activated
Umklapp
scattering rate which should enter the microwave conductivity. 
  Is Fig. 3
sufficient to conclude 
that these assertions are correct?  In order to examine the
true low-$T$ asymptotic forms, we have displayed  in Fig. 4 the
same curves and data  as in Fig. 3 on a log-log plot.   
>From the insert to the figure, it is clear
that the low-$T$ behavior of the two experimental relaxation
rates is quite different, but although the thermal conductivity
data is roughly consistent with the $T^3$ result expected from the
spin-fluctuation theory, the microwave data appear to lie
considerably above the activated prediction at the lowest
temperatures, a behavior hidden by the linear plot in Fig. 3. 
In the main part of the figure, we show the raw experimental data with no 
constant subtracted, together with the inelastic rates added to
a constant $1/\tau_0$ determined by the experimental data at low
$T$.   

The most natural interpretation of the residual scattering rates $1/\tau_0$
is in terms of disorder.  It is striking that the 
 $1/\tau_0$ extracted from the thermal conductivity
measurement appears to be significantly larger than that extracted from
the microwave measurements, since both measurements were performed
on nominally equivalent $BaZrO_3$ crucible-grown samples.     
The most popular treatment of disorder in the cuprates is the 
self-consistent t-matrix approximation (SCTMA) describing
strong scattering by
in-plane near-unitarity limit defects.  This theory predicts 
a strong energy dependence of the low-temperature elastic
scattering rate, however, inconsistent with the roughly 
constant behavior observed here\cite{QSB94}. 
 This difficulty was remedied
to some extent in Ref. \cite{HH00} by considering additional
scattering from spatial fluctuations of the order parameter.
Even in this approach, however, bare scatterers are considered to
be isotropic, leading to a vanishing of impurity vertex corrections
in local quantities; this implies the same relaxation time
for both thermal and microwave conductivities.  If the isotropic
scattering assumption is relaxed, however, as considered
by Durst and Lee\cite{DL99}, the $\omega\rightarrow 0$ microwave conductivity is found
to be strongly enhanced relative to the thermal conductivity,
which is unrenormalized.   This may be interpreted as a reduction
of the scattering rate in the microwave case, but a full 
frequency-dependent calculation and proof that a Lorentzian
conductivity obtains in analogy to Ref. \cite{HPS94} is required before one
may conclude that the observed discrepancy is caused
by anisotropic impurity scattering.

To summarize, using a d-wave BCS framework, we have
taken a simple spin-fluctuation interaction, 
which has been used to describe the
temperature dependence of the NMR $T_1$ and $T_2$ spin relaxation
times in YBCO below $T_c$, and have shown that it can also describe the
temperature dependence of both the inelastic  
charge and heat transport times below
$T_c$.  The contribution to the quasi-particle lifetime of normal 
spin-fluctuation scattering processes, which can relax  heat currents  but
not  charge currents, as well as Umklapp processes, have been calculated
separately. 
The difference in the two contributions was then shown to account
for the difference in  experimentally extracted lifetimes in 
recent microwave and heat conductivity
measurements on ultraclean samples of YBCO.  
Both the predicted temperature dependences in a spin-fluctuation
mediated interaction model,  the Umklapp  scattering 
rates $1/\tau_U\sim \exp -\Delta_U/T$
and  the total scattering rate 
$1/\tau\sim T^3$, appear to agree with the data, as do
the predicted absolute magnitudes of the two rates.  The exponential
form of $1/\tau_U$  implies a  sensitivity to the Fermi
surface shape, which is expected to lead to a significant
increase 
in $1/\tau_U$ at low temperatures upon underdoping.
  While our
results depend upon the band structure and interaction parameters, we have
selected these to be consistent with 
Fermi surface data and NMR measurements, and conclude that the results
for the microwave and thermal conductivity inelastic 
quasi-particle lifetimes are consistent with an effective electron-electron
interaction which is mediated by spin-fluctuations.


\acknowledgments

We thank D.A.~Bonn, S.~Daul, A.C.~Durst, P.A.~Lee,  
A.~Millis,  M.~Reizer,
and A.~Yashenkin for useful discussions, as
well as S.~Quinlan for the use of his code.  DD and
DJS acknowledge support under NSF DMR98-17242 and PJH acknowledges support 
under NSF DMR 99-74396.  DJS and PJH are grateful to the Institute for
Theoretical Physics for hospitality and support through NSF PHY94-07194 
during the completion of the manuscript.

\vfil

\newpage

\begin{figure}
\centerline{\epsfysize=6cm\epsfbox{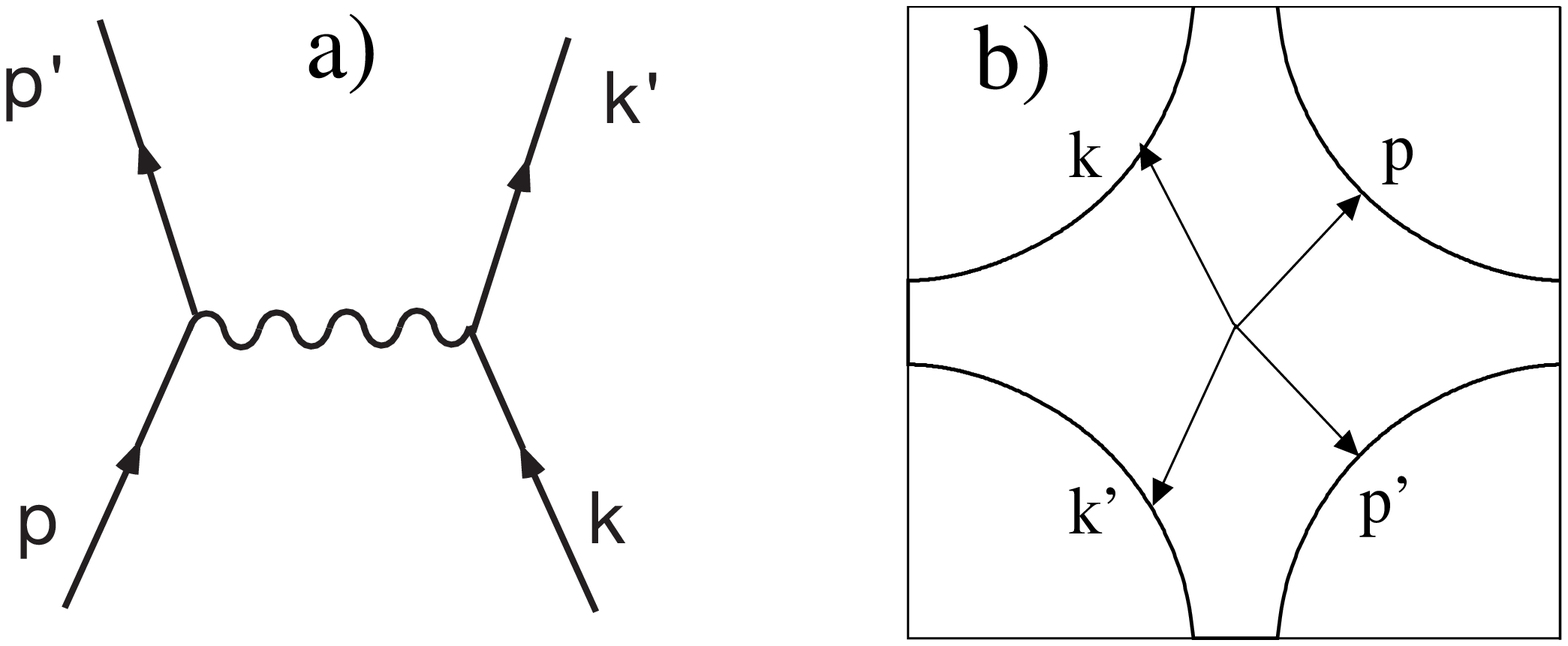}}
\vspace{0.5cm}
\caption{ (a) A schematic representation of the 
scattering process.
(b) The non-interacting Fermi surface 
with $t'/t=-0.35$ at a filling of $\en=0.85$.  The vectors
shown correspond to an Umklapp process in which $\bar p + \bar k = \vec
p^\prime + \vec k^\prime + (0, 2\pi/a)$.  
}
\label{fdiagram}
\end{figure}

\begin{figure}
\centerline{\epsfysize=7cm \epsfbox{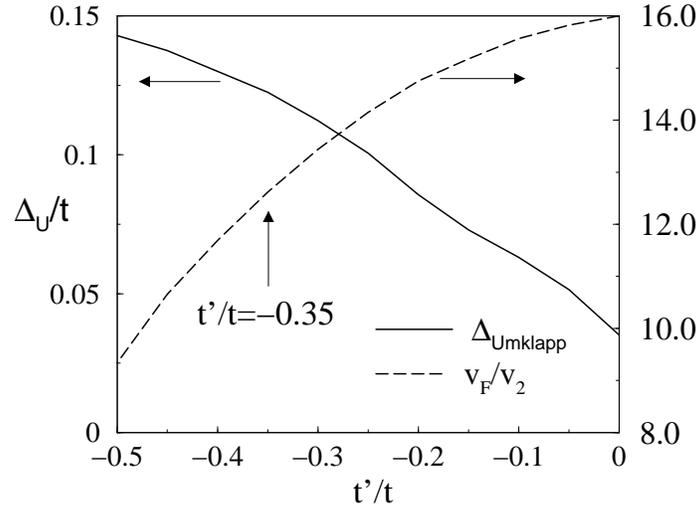}}
\vspace{0.5cm}
\caption{The size of the Umklapp gap $\Delta_U/t$ (solid curve) and the
$v_F/v_2$ ratio (dashed curve) versus the ratio of the next-near-neighbor
hopping $t^\prime$ to the near-neighbor hopping $t$.}
\label{fig2}
\end{figure}

\begin{figure}
\begin{picture}(150,300)
\leavevmode\centering\includegraphics{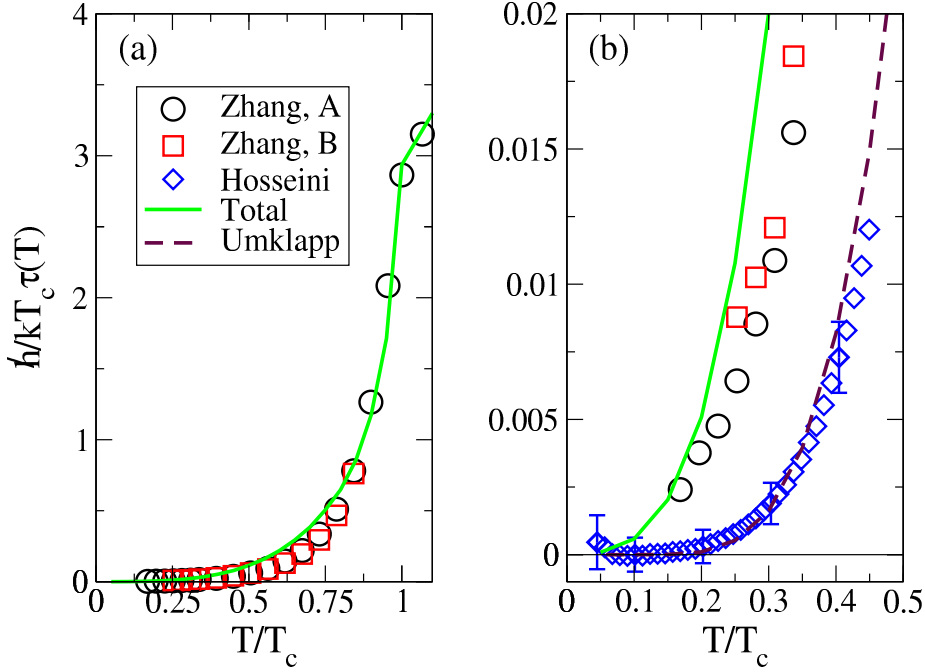}
\end{picture}

\vspace{0.5cm}
\caption{(a) The normalized YBCO quasi-particle scattering rates
$\hbar/(\tau k_B T_c)$ and $\hbar/(\tau_U k_B T_c)$ versus reduced
temperature $T/T_c$.  The experimental points for $\hbar/(\tau k_B
T_c)$ were obtained from the thermal conductivity mean-free path
measurements by Zhang, \etal~\protect\cite{Zhapp} using $v_F=3\times
10^7$ cm/sec and are shown as the open circles and squares. The solid
line is the total quasi-particle lifetime obtained from
Eq.~(\protect\ref{eq:tau}) on a $512\times 512$ lattice with
$\ubar/t=2.2, T_c=0.1t, t'/t=-0.35, \2del=5$, and $\en=0.85$. (b) Same as
in (a) for reduced temperatures less than 0.5 with the data of
Hosseini, \etal~\protect\cite{Hos99} included as open diamonds.  The
solid line corresponds to the total lifetime calculated as in (a),
while the dashed line represents the lifetime due to Umklapp processes
only, Eq.~(\protect\ref{seven}).  Note a constant term has been
subtracted from the experimental data.} 
\label{fig3}
\end{figure}

\vskip -2cm
\begin{figure}[h]
\begin{picture}(150,300)
\leavevmode\centering\includegraphics{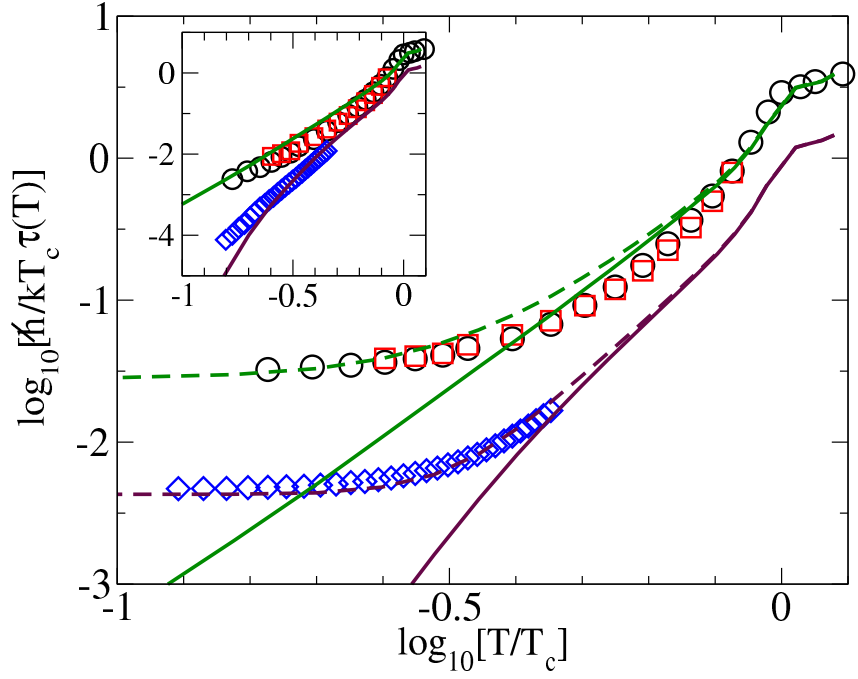}
\end{picture}
\caption{Log-log plot of theoretical predictions for normalized total
and Umklapp scattering rates $\hbar/(\tau k_B T_c)$ and $\hbar/(\tau_U
k_B T_c)$ obtained from Eq.~(\protect\ref{eq:tau}) (solid lines), same
parameters as Fig.~\protect\ref{fig3}.  Data from
Refs.~\protect\cite{Zhapp} and ~\protect\cite{Hos99} as in
Fig.~\protect\ref{fig3}.  Dashed curves: same theoretical predictions
for $\hbar/(\tau k_B T_c)$ and $\hbar/(\tau_U k_B T_c)$ with constants
$1/\tau_0= 5.0 \times 10^{10}$ s$^{-1}$ and 3.3 $\times 10^{11}$
s$^{-1}$ added to $1/\tau_U$ and $1/\tau$, respectively.  Insert: same
as main panel but with $1/\tau_0$ subtracted from data.}

\label{fig4}
\end{figure}



\end{document}